\documentstyle[11pt]{article}
\begin{document}
\begin{titlepage}

\hfill{BONN-HE-92-16}\\
\hfill{\em revised version}
\vskip.3in
\begin{center}
{\huge Superconformal Affine Liouville Theory}
\vskip.3in
{\Large Francesco Toppan}\footnote{Address after October 1, 1992:
Laboratoire de Physique Th\'{e}orique ENSLAPP, Ecole
Normale Sup\'erieure de Lyon, 46 All\'ee d'Italie,
69007 Lyon, France}
and {\Large Yao-Zhong Zhang}
\footnote{Address after 28 October 1992: Department of Mathematics,
University of Queensland, Brisbane, Qld 4072, Australia}
\vskip.3in
Physikalisches Institut, Universit\"at Bonn, Nussallee 12,
      D-5300 Bonn 1, Germany
\end{center}
\vskip 3cm
\begin{center}
{\bf Abstract}
\end{center}
We present a superconformally invariant
and integrable model based on the twisted affine Kac-Moody superalgebra
$\hat{osp(2|2)}^{(2)}$ which
is the supersymmetrization of the purely bosonic conformal affine Liouville
theory recently proposed by Babelon and Bonora. Our model
reduces  to the super-Liouville or to the super
sinh-Gordon theories under certain limit conditions and can be
obtained, via
hamiltonian reduction, from  a superspace WZNW model with values in the
corresponding affine KM supergroup. The reconstruction formulae for classical
solutions are given. The classical $r$-matrices in the homogeneous grading
and the exchange algebras are worked out.
\end{titlepage}

\section{Introduction}

Associated to a given bosonic Lie algebra, three kinds
of models can be introduced
in terms of a Lax pair formalism: they are
related to the algebra itself,
to its loop extension and to its affine Kac-Moody (KM) extension, respectively.
For instance in the case of $sl(2)$
the three models are the following:
i) the Liouville theory,  which is conformally
invariant and integrable;
ii) the sinh-Gordon (SG) theory (based on the loop algebra  $\tilde{sl(2)}$)
which is still integrable but no longer conformally invariant;
iii) the conformal affine Liouville (CAL) model proposed by Babelon and Bonora
\cite{BB1} (based on the affine KM algebra $\hat{sl(2)}$), which is
like the Liouville theory conformally invariant and integrable;
moreover
the Liouville and the SG theories
can be recovered from it with two different limit procedures.

Similar steps can be performed in correspondence with any Toda and
WZNW-theory \cite{BMZ1}.

It is natural to look for supersymmetric generalizations of
the above construction. The
supersymmetric extensions of the Liouville and SG theories
have indeed been extensively studied \cite{Arvis}\cite{LSL et al}
\cite{Olshanesky}.
The super-Liouville theory, which is superconformally invariant
and integrable, is found to correspond to an obvious
superalgebra generalization - $osp(1|2)$ - of the
bosonic algebra $sl(2)$.
However, as explicit computations show,
the counterparts of the bosonic construction based on the
loop superalgebra $\tilde{osp(1|2)}$ and the affine KM superalgebra
$\hat{osp(1|2)}$
do not work. Rather it turns out that the superalgebra
underlying the integrable but no longer superconformally invariant
super-SG theory is the twisted loop superalgebra
 $\tilde{osp(2|2)^{(2)}}$.
Therefore one expects
that a supersymmetric counterpart of the bosonic CAL-theory
should arise from the construction based on the twisted affine KM
superalgebra $\hat{osp(2|2)^{(2)}}$. This is indeed the case.
In this paper we generalize the Babelon-Bonora construction to the
$N=1$
supersymmetric case and provide a new
superconformally invariant and integrable
model (let us call it super-CAL theory).
Our super-CAL model
realizes an interpolation between super-Liouville an super-SG theories
which are recovered as special limits. Moreover its
bosonic sector corresponds to the
CAL theory.
As in the bosonic case
\cite{B...W}\cite{A...Z}\cite{BMZ2}, the super-CAL theory can be obtained, via
hamiltonian reduction, from the superspace WZNW model with values
in the corresponding affine KM supergroup.
We will work at any step
with a manifestly supersymmetric formalism.

\section {Notation: Twisted Affine KM Superalgebra $\hat{osp(2|2)}^{(2)}$}
This section is devoted to a brief account for the twisted affine KM
superalgebra $\hat{osp(2|2)}^{(2)}$ \cite{Frappat et al} and to introduce
some useful notations.

The superalgebra $osp(2|2)$ is a ${\bf Z}_2$ graded algebra, $osp(2|2)=
osp(2|2)^{\rm even}\oplus osp(2|2)^{\rm odd}$, with
\begin{eqnarray}
&&osp(2|2)^{\rm even}=u(1)\oplus sl(2)=\{H'\}\oplus \{H,E_\alpha,
   E_{-\alpha}\}\nonumber\\
&&osp(2|2)^{\rm odd}=\{F_\beta, F_{-\beta}, F_{\bar{\beta}},
   F_{-\bar{\beta}}\}
\end{eqnarray}
These generators satisfy the following (anti-)commutation relations,
\begin{eqnarray}
&&\{F_{\pm\beta}, F_{\pm\beta}\}=0=\{F_{\pm\bar{\beta}},
  F_{\pm\bar{\beta}}\},~~~~\{F_{\pm\beta}, F_{\pm\bar{\beta}}\}=\pm
  E_{\pm\alpha}\nonumber\\
&&\{F_\beta, F_{-\beta}\}=-\frac{1}{2}(H-H'),~~~~~\{F_{\bar{\beta}},
  F_{-\bar{\beta}}\}=-\frac{1}{2}(H+H')\nonumber\\
&&[E_{\pm\alpha}, F_{\mp\beta}]=F_{\pm\bar{\beta}},~~~~~~[E_{\pm\alpha},
  F_{\mp\bar{\beta}}]=F_{\pm\beta}\nonumber\\
&&[E_{\pm\alpha}, F_{\pm\beta}]=0=[E_{\pm\alpha}, F_{\pm\bar{\beta}}],~~~~[H,
  F_{\pm\beta}]=\pm F_{\pm\beta}\nonumber\\
&&[H, F_{\pm\bar{\beta}}]=\pm F_{\pm\bar{\beta}},~~~~~~[H, E_{\pm \alpha}]
  =\pm 2E_{\pm \alpha}\nonumber\\
&&[E_\alpha, E_{-\alpha}]=H, ~~~~~~~[H', E_{\pm\alpha}]=0\nonumber\\
&&[H', F_{\pm\beta}]=\pm F_{\pm\beta},~~~~~~[H',
  F_{\pm\bar{\beta}}]=\mp F_{\pm\bar{\beta}}
\end{eqnarray}
The fundamental representation of $osp(2|2)$ is 3-dimensional,
\begin{eqnarray}
&&H=e_{11}-e_{22},~~~~E_\alpha=e_{12},~~~~E_{-\alpha}=e_{21},~~~~H'=
  e_{11}+e_{22}+2e_{33}\nonumber\\
&&F_\beta=e_{32},~~~~F_{-\beta}=e_{23},~~~~F_{\bar{\beta}}=e_{13},~~~~
  F_{-\bar{\beta}}=-e_{31}
\end{eqnarray}
with $e_{ij}$ being $3\times 3$ supermatrix and
$(e_{ij})_{kl}=\delta_{ik}\delta_{jl}$.
The nondegenerate, invariant and supersymmetric bilinear form on $osp(2|2)$
is simply the usual supertrace "str" which is defined as
\begin{equation}
{\rm str} X={\rm tr}a-{\rm tr}b,~~~~~{\rm for}~~X=\left (
\begin{array}{cc}
a & c\\
d & b
\end{array}
\right )\label{Killing}
\end{equation}
The superalgebra $osp(2|2)$ has an automorphism $\tau$ of order 2 and
w.r.t. the eigenvectors of $\tau$ we have the decomposition,
$osp(2|2)=osp(2|2)_0\oplus osp(2|2)_1$, where,
\begin{eqnarray}
&&osp(2|2)_0\equiv \{X\in osp(2|2),~\tau (X)=X\}=\{H, E_\alpha, E_{-\alpha},
  i(F_\beta+F_{\bar{\beta}}), i(F_{-\beta}+F_{-\bar{\beta}})\}\nonumber\\
&&osp(2|2)_1\equiv \{X\in osp(2|2),~\tau (X)=-X\}=\{H',
  (F_\beta-F_{\bar{\beta}}), (F_{-\beta}-F_{-\bar{\beta}})\}\label{osp01}
\end{eqnarray}
The twisted loop superalgebra $\tilde{osp(2|2)}^{(2)}$ is defined  as
\cite{Olshanesky}
\begin{eqnarray}
\tilde{osp(2|2)}^{(2)}&=&\sum_{n\in  {\bf Z}}\lambda^n
    osp(2|2)_{n\;{\rm mod}2}\nonumber\\
   &=&\{\lambda^{2n}H, \lambda^{2n}E_{\pm\alpha},i\lambda^{2n}(F_{\pm\beta}
  +F_{\pm\bar{\beta}}),\lambda^{2n+1}H', \lambda^{2n+1}(F_{\mp\beta}
   -F_{\mp\bar{\beta}}),~ n\in {\bf Z}\}\label{loop}
\end{eqnarray}
where the spectral parameter $\lambda$ represents the homogeneous grading.
We will always work with the homogeneous grading throughout this paper.

The twisted affine KM superalgebra $\hat{osp(2|2)}^{(2)}$ is obtained by adding
to the twisted loop superalgebra $\tilde{osp(2|2)}^{(2)}$ the center $\hat{c}$
and the derivative $\hat{d}=\lambda\frac{d}{d\lambda}$. Namely,
 $\hat{osp(2|2)}^{(2)}=
\tilde{osp(2|2)}^{(2)}\oplus {\bf C}\hat{c}\oplus {\bf C}\hat{d}$. The
graded bracket on $\hat{osp(2|2)}^{(2)}$ reads
\begin{equation}
[\hat{X}, \hat{Y}\}_\wedge=[\tilde{X}, \tilde{Y}\}_\sim+\hat{c}\frac{1}{2\pi i}
  \oint\;d\lambda\;{\rm Str}[(\partial_\lambda\tilde{X}(\lambda))
  \tilde{Y}(\lambda)]
\end{equation}
where "Str" stands for the invariant, nondegenerate and supersymmetric
bilinear form on $\hat{osp(2|2)}^{(2)}$ which is defined as
\begin{eqnarray}
&&{\rm Str}(\hat{X}^m_{\alpha_i}\hat{X}^n_{-\alpha_j})=\delta_{m+n,0}
  {\rm str}(X_{\alpha_i}X_{-\alpha_j}),~~~~{\rm Str}(\hat{H}^m\hat{H}^n)=
  \delta_{m+n,0}{\rm str}(HH)\nonumber\\
&&{\rm Str}(\hat{H'}^m\hat{H'}^n)=\delta_{m+n,0}{\rm str}(H'H'),~~~~{\rm Str}
  (\hat{c}\hat{d})=1,~~~~{\rm the~ rests}=0
\end{eqnarray}
where the quantities
$\hat{T}^m\equiv \lambda^mT,~ m=2{\bf Z}~ {\rm or}~2{\bf Z}+1$.

We perform the Cartan decomposition,
\begin{equation}
\hat{osp(2|2)}^{(2)}=\hat{\cal G}_<\oplus \hat{\cal G}_=\oplus
   \hat{\cal G}_>\label{decomposition1}
\end{equation}
where the Cartan subalgebra $\hat{\cal G}_=$ is spanned by
$\{H, \hat{c}, \hat{d}\}$, $\hat{\cal G}_>$ is the nilpotent subalgebra
generated by positive root vectors,
\begin{equation}
\lambda^{2n}H,E_\alpha, \lambda^{2n}E_{\pm\alpha}, F_{\alpha_1}, \lambda^{2n}
 F_{\pm\alpha_1}, F_{\alpha_2}, \lambda^{2n}F_{\alpha_2},
\lambda^{2n}F_{-\alpha_2}, \lambda H',\lambda^{2n+1}H',~~n>0
\end{equation}
and $\hat{\cal G}_<$ is the nilpotent subalgebra generated by negative root
vectors,
\begin{equation}
\lambda^{-2n}H,E_{-\alpha}, \lambda^{-2n}E_{\pm\alpha}, F_{-\alpha_1},
 \lambda^{-2n}
 F_{\pm\alpha_1}, \lambda^{-2n}F_{\alpha_2},F_{-\alpha_2},
\lambda^{-2n}F_{-\alpha_2}, \lambda^{-1}H', \lambda^{-(2n+1)}H',~~n>0
\end{equation}
In the above,
\begin{equation}
F_{\pm\alpha_1}\equiv i(F_{\pm\beta}+F_{\pm\bar{\beta}}),~~~~F_{\pm\alpha_2}
 \equiv \lambda^{\pm 1}(F_{\mp\beta}-F_{\mp\bar{\beta}})
\end{equation}
are odd simple root vectors of the twisted affine KM superalgebra
$\hat{osp(2|2)}^{(2)}$.

\section{Super-CAL: Zero-Curvature Formulation}
The super-CAL is a super-Toda system associated to the twisted affine
KM superalgebra $\hat{osp(2|2)}^{(2)}$. It is the generalization of the purely
bosonic CAL\cite{BB1} to the $N=1$ supersymmetric case.

Let $x^+, \theta^+, x^-$ and $\theta^-$ denote the light-cone coordinates of
the world-sheet (1,1) superspace. The superderivatives are
\begin{equation}
D_+=\frac{\partial}{\partial\theta^+}+i\theta^+\partial_+,~~~~~~D_-=
 \frac{\partial}{\partial\theta^-}+i\theta^-\partial_-
\end{equation}
which satisfy $(D_\pm)^2=i\partial_\pm$ and $\{D_+,D_-\}=0$.
We introduce the scalar superfield $\Psi(x^+,\theta^+,x^-,\theta^-)$ with
values in the Cartan subalgebra of $\hat{osp(2|2)}^{(2)}$,
\begin{equation}
\Psi={\textstyle\frac{1}{2}}\Phi H+a{\textstyle\frac{1}{2}}\Lambda\hat{d}+
{\textstyle\frac{1}{2}}\Xi\hat{c}
\end{equation}
where $\Phi, \Lambda$ and $\Xi$ are scalar superfields
and $a$ is a real parameter.
The Ramond or Neveu-Schwarz boundary conditions will be assumed for
fermionic fields.
We define two Lax-pair superpotentials,
\begin{equation}
{\cal L}_+=D_+\Psi+e^{{\rm ad}\Psi}{\cal E}_+,~~~~{\cal L}_-=-D_-\Psi+
e^{-{\rm ad}\Psi}{\cal E}_-
\end{equation}
where ${\cal E}_+$ and ${\cal E}_-$ are respectively the sum of odd positive
and
negative simple root vectors,
\begin{equation}
{\cal E}_+=F_{\alpha_1}+F_{\alpha_2},~~~~~~{\cal E}_-=F_{-\alpha_1}+
  F_{-\alpha_2}
\end{equation}
and write the linear system
\begin{equation}
(D_\pm+{\cal L}_\pm){\cal T}=0
\end{equation}
where ${\cal T}$ belongs to the affine KM supergroup whose Lie algebra is
$\hat{osp(2|2)}^{(2)}$.

The zero-curvature condition, which is the compatibility condition of the
above linear system,
\begin{equation}
D_+{\cal L}_-+D_-{\cal L}_++\{{\cal L}_+,{\cal L}_-\}=0
\end{equation}
can be worked out using only the Lie superalgebra strucutre of
$\hat{osp(2|2)}^{(2)}$. We get
\begin{eqnarray}
&&D_+D_-\Phi=e^\Phi-e^{a \Lambda -\Phi}\nonumber\\
&&D_+D_-\Lambda =0\nonumber\\
&&D_+D_-\Xi=e^{a \Lambda -\Phi}\label{super-CAT1}
\end{eqnarray}
which define our super-CAL.

{\bf Remarks}:

(i) The super-CAL is superconformally invariant; this is seen as
follows. Let $Z^+=(x^+,\theta^+)$ and
$Z^-=(x^-,\theta^-)$. Then the superconformal symmetry of the super-CAL is
realized as
\begin{eqnarray}
&&Z^+\longrightarrow \tilde{Z}^+(Z^+)=(\tilde{x}^+(Z^+),\tilde{\theta}^+
  (Z^+))\,,~~~~Z^-\longrightarrow \tilde{Z}^-(Z^-)=(\tilde{x}^-(Z^-),
  \tilde{\theta}^-(Z^-))\nonumber\\
&&\Phi(Z^+,Z^-)\longrightarrow \tilde{\Phi}(\tilde{Z}^+,\tilde{Z}^-)=
  \Phi(\tilde{Z}^+,\tilde{Z}^-)+\ln (D_+\tilde{\theta}^+
   D_-\tilde{\theta}^-)\nonumber\\
&&\Lambda (Z^+,Z^-)\longrightarrow \tilde{\Lambda }(\tilde{Z}^+,\tilde{Z}^-)=
  \Lambda (\tilde{Z}^+,\tilde{Z}^-)+{2\over a}\ln (D_+\tilde{\theta}^+
   D_-\tilde{\theta}^-)\nonumber\\
&&\Xi(Z^+,Z^-)\longrightarrow \tilde{\Xi}(\tilde{Z}^+,\tilde{Z}^-)=
  \Xi(\tilde{Z}^+,\tilde{Z}^-)-{\textstyle\frac{1}{2}}B\ln (D_+\tilde{\theta}^+
   D_-\tilde{\theta}^-)\label{conf. trans.}
\end{eqnarray}
where $B$ is an arbitrary constant. It is easy to show that (\ref{super-CAT1})
is indeed invariant under the transformations (\ref{conf. trans.}) thanks to
the covariant transformation property of $D_\pm$ under
$Z^\pm\rightarrow\tilde{Z}^\pm (Z^\pm)$ \cite{Gaume},
\begin{equation}
D_\pm =(D_\pm\tilde{\theta}^\pm )\tilde{D}_\pm\,,~~~~\tilde{D}_\pm=
\frac{\partial}{\partial\tilde{\theta}^\pm}+i\tilde{\theta}^\pm
\frac{\partial}{\partial \tilde{x}^\pm}
\end{equation}
which is subject to the constraints
\begin{equation}
D_\pm \tilde{x}^\pm=\tilde{\theta}^\pm D_\pm\tilde{\theta}^\pm
\end{equation}

(ii) In (\ref{super-CAT1}) the equation for the
superfield $\Phi$ reduces to the equations of motions
for the super-Liouville or for the super-SG theories
under the limits $a \rightarrow-\infty$
and  $a\rightarrow 0$, respectively;
therefore the super-CAL
provides an interpolation between the super-Liouville and the super SG.
{}From now on we work with the fixed normalization $a=1$.

We now expand the scalar superfields in terms of their components,
$\Phi=\phi-i\theta^+\psi_++i\theta^-\psi_-+i\theta^+\theta^-f$,
$\Lambda=\eta-i\theta^+\alpha_++i\theta^-\alpha_-+i\theta^+\theta^-g
$ and $\Xi=\xi-i\theta^+\beta_++i\theta^-\beta_-+i\theta^+\theta^-h$.
Eliminating the auxiliary fields $f,g$ and $h$ in the
super-CAL (\ref{super-CAT1})
($f=i(e^\phi-e^{\eta-\phi}), h=ie^{\eta-\phi}, g=0$), we are left with,
in components,
\begin{eqnarray}
&&\partial_+\partial_-\phi=e^{2\phi}-e^{2\eta-2\phi}+\psi_+\psi_- e^\phi
  -(\alpha_+-\psi_+)(\alpha_--\psi_-)e^{\eta-\phi}\nonumber\\
&&\partial_+\partial_-\eta=0,~~~~\partial_+\alpha_-=0=\partial_-
  \alpha_+\nonumber\\
&&\partial_+\psi_-=-i(\psi_+e^\phi+(\alpha_+-\psi_+)e^{\eta-\phi})\nonumber\\
&&\partial_-\psi_+=i(\psi_-e^\phi+(\alpha_--\psi_-)e^{\eta-\phi})\nonumber\\
&&\partial_+\partial_-\xi=e^{2\eta-2\phi}-e^\eta+(\alpha_+-\psi_+)(\alpha_--
  \psi_-)e^{\eta-\phi}\nonumber\\
&&\partial_+\beta_-=i(\alpha_+-\psi_+)e^{\eta-\phi},~~~~\partial_-\beta_+
  =-i(\alpha_--\psi_-)e^{\eta-\phi}
\end{eqnarray}
Setting the fermionic components $\psi_\pm, \alpha_\pm$ and $\beta_\pm$ to
be zero, we get the purely bosonic limit of the super-CAL,
\begin{eqnarray}
&&\partial_+\partial_-\phi=e^{2\phi}-e^{2\eta-2\phi}\nonumber\\
&&\partial_+\partial_-\eta=0\nonumber\\
&&\partial_+\partial_-\xi=e^{2\eta-2\phi}-e^\eta
\end{eqnarray}
We thus recognize the purely bosonic CAL \cite{BB1},
by the following redefinition of the $\xi$ field
\begin{equation}
\xi\rightarrow\xi-\int^{x^+}\,d{x^+}'\,\int^{x^-}\,d{x^-}'\;e^{\eta({x^+}',
 {x^-}')}
\end{equation}
It follows that the super-CAL is the supersymmetrization of the purely
bosonic CAL.

Notice that the super-CAL can be obtained from the action,
\begin{equation}
S=\int dx^+dx^-d\theta^+d\theta^-\;\left( D_+\Phi D_-\Phi+D_+\Xi D_-\Lambda
  +D_+\Lambda D_-\Xi+2e^\Phi+2e^{\Lambda -\Phi}\right )\label{s-action}
\end{equation}

\section {Super-CAL as Constrained super-WZNW}
In this section we show that the super-CAL can be obtained from
superspace WZNW reduction
\cite{Sorba} by implementing certain constraints in the left
and right KM supercurrents simultaneously.

A remarkable difference
between the superspace supergroup-valued WZNW reduction
studied here and the non-superspace supergroup-valued WZNW reduction
considered in
\cite{Inami} is
that in the former case one has a whole set of first-class
constraints and in the latter case a whole system of first-class constraints
can be chosen only if auxiliary fields are introduced into the constraints.

Let $G(x^+,\theta^+,x^-,\theta^-)$ be a superfield with values in the
affine KM supergroup whose Lie algebra is $\hat{osp(2|2)}^{(2)}$, formally
written
as ${\rm exp} (\hat{osp(2|2)}^{(2)})$. We start with the manifestly
supersymmetric WZNW action with values in the affine KM supergroup
${\rm exp}(\hat{osp(2|2)}^{(2)}$,
\begin{eqnarray}
\kappa S_{\rm WZNW}(G)&=&\frac{\kappa}{2}\int\,d^2xd^2\theta\;\left
   [{\rm Str}(G^{-1}D_+GG^{-1}D_-G)\right .\nonumber\\
   & &\left .+\int\,dt\;{\rm Str}(G^{-1}\partial_tG\{G^{-1}D_+G,G^{-1}D_-G\}
  )\right ]\label{s-wznw}
\end{eqnarray}
The corresponding equations of motion read
\begin{equation}
D_+{\cal J}=0,~~~~~~~D_-\bar{\cal J}=0\label{eq. of wznw}
\end{equation}
where
\begin{equation}
{\cal J}=G^{-1}D_-G,~~~~~~~~\bar{\cal J}=-D_+GG^{-1}
\end{equation}
are left and right KM supercurrents taking values in $\hat{osp(2|2)}^{(2)}$.

With respect to the Cartan decomposition formula (\ref{decomposition1}) we
can write a general supergroup element in the following form
\begin{equation}
G=G_<G_=G_>\label{gauss}
\end{equation}
where $G_=={\rm exp}[-(\Phi H+\Lambda\hat{d}+\Xi\hat{c})]$ and $G_<\in
{\rm exp}{\hat{\cal G}_<}, G_>\in {\rm exp} \hat{\cal G}_>$. Correspondingly,
the supercurrents ${\cal J}$ and $\bar{\cal J}$ are separated into three
parts, respectively,
\begin{equation}
{\cal J}={\cal J}_<+{\cal J}_=+{\cal J}_>,~~~~~\bar{\cal J}=\bar{\cal J}_<
  +\bar{\cal J}_=+\bar{\cal J}_>
\end{equation}
Inserting (\ref{gauss}) into the WZNW equations of motion (\ref{eq. of wznw}),
we can show that
(\ref{eq. of wznw}) becomes the "zero-curvature conditions"
\begin{eqnarray}
&&D_+(G^{-1}_=J_<G_=+G^{-1}_=D_-G_=)+D_-\bar{J}_>+\{\bar{J}_>,G_=^{-1}J_<G_=+
  G^{-1}_=D_-G_=\}=0\nonumber\\
&&D_-(G_=\bar{J}_>G^{-1}_=+D_+G_=G^{-1}_=)+D_+J_<+\{G_=\bar{J}_>G^{-1}_=+D_+
  G_=G^{-1}_=, J_<\}=0\label{zero-curvature}
\end{eqnarray}
where $J_<=G^{-1}_<D_-G_<$ and $\bar{J}_>=-D_+G_>G^{-1}_>$, respectively
belonging to $\hat{\cal G}_<$ and $\hat{\cal G}_>$.

We introduce constant elements
$M_<\in \hat{\cal G}_<$, $\bar{M}_>\in \hat{\cal G}_>$,
which are Grassmann odd and have non-zero components along only the simple
roots. Namely,
\begin{equation}
M_<=\mu_<^1F_{-\alpha_1}+\mu_<^2F_{-\alpha_2},~~~~~~\bar{M}_>=\mu_>^1
F_{\alpha_1}+\mu_>^2F_{\alpha_2}
\end{equation}
where $\mu^1_<, \mu^2_<, \mu_>^1$ and $\mu^2_>$ are non-zero real constants.

Let us impose the constraints
\begin{equation}
J_<=G_=M_<G_=^{-1},~~~~~~\bar{J}_>=-G^{-1}_=\bar{M}_>G_=\label{constraint1}
\end{equation}
then some algebraic works show that the eq.(\ref{zero-curvature}) takes the
form
\begin{eqnarray}
&&D_+D_-\Phi=-\mu_>^1\mu_<^1e^\Phi+\mu_>^2\mu_<^2e^{\Pi-\Phi}\nonumber\\
&&D_+D_-\Lambda =0\nonumber\\
&&D_+D_-\Xi=-\mu_>^2\mu_<^2e^{\Pi-\Phi}\label{super CAT2}
\end{eqnarray}
which are nothing but the super-CAL equations (\ref{super-CAT1}) if we set
$\mu_>^1\mu_<^1=-1=\mu_>^2\mu_<^2$.

It can be shown that the constraints (\ref{constraint1}) are equivalent to
\begin{equation}
{\cal J}_<=(G^{-1}_>M_<G_>)_>=M_<,~~~~~\bar{\cal
J}_>=-(G_<\bar{M}_>G^{-1}_<)_>=
 -\bar{M}_>\label{constraint2}
\end{equation}
thanks to the fact that the $M_<$ and $\bar{M}_>$ consist of only odd simple
root vectors. Eq.(\ref{constraint2}) implies that (\ref{constraint1}) can be
expressed as linear constraints on the left and right affine KM supercurrents.

We now present a Lagrangian realization of the above WZNW reduction. For this
purpose we need to impose the constraints of the type (\ref{constraint2}) by
means of Lagrange multipliers. This can be implemented by gauging the
superspace WZNW action (\ref{s-wznw}) as follows. Introduce the fermionic
gauge superfields $A_<$ and $A_>$ which take values in the nilpotent
subalgebra $\hat{\cal G}_<$ and $\hat{\cal G}_>$, respectively. The gauge
invariant action turns out to be
\begin{eqnarray}
\kappa S_{\rm gauged}(G,A_<,A_>)&=&\kappa S_{\rm WZNW}(G)+\kappa\int\,
 d^2xd^2\theta\;{\rm Str}[A_>(G^{-1}D_-G-M_<)\nonumber\\
  & &+A_<(D_+GG^{-1}-\bar{M}_>)+A_<GA_>G^{-1}]\label{s-gauged}
\end{eqnarray}
One can easily check the invariance of this action under the following super
gauge transformations,
\begin{equation}
G\rightarrow g_<Gg^{-1}_>,~~~~A_>\rightarrow g_>A_>g^{-1}_>+D_+g_>g^{-1}_>,
{}~~~~~~A_<\rightarrow g_<A_<g^{-1}_<-D_-g_<g^{-1}_<
\end{equation}
where the superfield $g_<\in {\rm exp}\hat{\cal G}_<$ and the superfield
$g_>\in {\rm exp}\hat{\cal G}_>$.

In the gauge $A_<=0=A_>$, the Euler-Lagrange equations derived from the gauged
action reproduce the equations of motion together with the constraint
(\ref{constraint2}). Therefore the gauged WZNW model is entirely equivalent to
the constrained WZNW theory.

On the other hand, there exists the diagonal gauge, $A_>=G^{-1}_=\bar{M}_>G_=$
and $A_<=G_=M_<G^{-1}_=$. In this gauge the constraints set $G$ in the maximal
torus generated by the Cartan subalgebra $\hat{\cal G}_=$, that is, $G=G_=$.
Taking these facts into account and we have, from the action (\ref{s-gauged}),
\begin{equation}
S_{\rm eff}(G_=)=S_{\rm WZNW}(G_=)-\int\,d^2xd^2\theta\;{\rm Str}
 (G_=M_<G^{-1}_=\bar{M}_>)\label{s-eff}
\end{equation}
which coincides with the action (\ref{s-action}) of the super-CAL if we set
$\mu_<^1\mu_>^1=-1=\mu_<^2\mu_>^2$.

\section{The Classical Solutions}
In this section, we give a Leznov-Saveliev reconstruction theorem\cite{Leznov}
for the classical solution to the super-CAL.

Instead of working with the transfer matrix ${\cal T}$
it is convenient to introduce the matrices $Q, {\overline Q}$
belonging to the affine KM supergroup, through the relations
\begin{equation}
Q=e^{-{ \Psi}} {\cal T},~~~~~~{\overline Q}= {\cal T}^{-1}
e^{-{\Psi}}\label{add.1}
\end{equation}
Consider now any representation of the superalgebra
with highest weight vector $|\lambda^{(r)}_{\rm max}>$, namely
${\cal E}_+|\lambda^{(r)}_{\rm max}>=0$,
$<\lambda^{(r)}_{\rm max}|{\cal E}_-=0$.
then, once the superfields
$\xi^{(r)} =<\lambda^{(r)}_{\rm max}|Q $ and
$\overline{\xi}^{(r)}=\overline{Q}|\lambda^{(r)}_{\rm max}>$ are introduced,
they turn out to be superchiral:
$ D_+{\overline \xi}^{(r)}=D_- \xi^{(r)} =0 $,
namely $\bar{\xi}^{(r)}=\bar{\xi}^{(r)}(x^-,\theta^-)$ and
$\xi^{(r)}=\xi^{(r)}(x^+,\theta^+) $.
The whole dynamics of the super-CAL is contained
in the fields $\xi^{(r)}$ and
$\overline{\xi}^{(r)}$. The reconstruction formula expressing the original
superfields $\Phi,\Lambda$ and $\Xi$ in terms of $\xi^{(r)}$ and
$\overline{\xi}^{(r)}$ is
\begin{equation}
e^{-2\lambda^{(r)}_{\rm max}(\Psi)}=\xi^{(r)}
\overline{\xi}^{(r)}\label{construction1}
\end{equation}
with $\lambda^{(r)}_{\rm max}$ being highest weight corresponding to the
highest weight state $|\lambda^{(r)}_{\rm max}>$.

We can perform as in the ordinary case the Gauss
decomposition of $Q$ and $\overline{Q}$:
let us write
$Q=N_<e^{K_=}N_>$, ${\overline Q}={\overline N}_<e^{{\overline K}_=}
{\overline N}_>$,
where $K_=,{\overline K}_=$ live in the Cartan subalgebra,
$N_>,{\overline N}_>~\in {\rm exp}\hat{\cal G}_>$ and
$N_<,{\overline N}_<~\in {\rm exp}\hat{\cal G}_<$ are respectively strictly
upper and lower triangular super matrices.
As a consequence of the linear systems satisfied by $Q$ and ${\overline Q}$,
the following equations hold
\begin{eqnarray}
&&D_-K_==D_-N_>=0,~~~~D_+N_> {N_>}^{-1}=- e^{-{\rm ad} K_=}
  {\cal E}_+\nonumber\\
&&D_+\overline{K}_==D_+\overline{N}_<=0 ,~~~~\overline{N}_<^{-1} D_-
\overline{N}_<=e^{{\rm ad} \overline{K}_=}{\cal E}_-
\end{eqnarray}
which tell us that $N_>,\overline{N}_<, K_=$ and $\overline{K}_=$ are chiral
superfields and $N_>,\overline{N}_<$ can be solved in terms
of $K_=$ and $\overline{K}_=$  respectively.
The reconstruction formula can be expressed as
\begin{equation}
\xi^{(r)}\overline{\xi}^{(r)}=
<\lambda^{(r)}_{\rm max}|
e^{K_=}N_>{\overline N}_<e^{{\overline K}_=}
|\lambda^{(r)}_{\rm max}>\label{reconstruction2}
\end{equation}

This reconstruction formula can also be obtained via hamiltonian reduction
from the classical solutions of the super-WZNW model.

The classical equation of motion (\ref{eq. of wznw}) implies
\begin{equation}
G(x^+,\theta^+,x^-,\theta^-)=U(x^+,\theta^+)V(x^-,\theta^-)\label{solution1}
\end{equation}
where the chiral superfields $U(x^+,\theta^+)$ and $V(x^-,\theta^-)$ belong
to the affine KM supergroup. Applying the decomposition formula for $G, U$
and $V$, (\ref{solution1}) takes the form
\begin{eqnarray}
&&G_<(x^+,\theta^+,x^-,\theta^-)G_=(x^+,\theta^+,x^-,\theta^-)G_>(x^+,
\theta^+,x^-,\theta^-)\nonumber\\
&&=U_<(x^+,\theta^+)e^{K_+(x^+,\theta^+)}U_>(x^+,\theta^+)
V_<(x^-,\theta^-)e^{K_-(x^-,\theta^-)}V_>(x^-,\theta^-)\label{1}
\end{eqnarray}
with
$K_+=\Theta_+ H-\Lambda_+\hat{d}+\Upsilon_+\hat{c}$,
$K_-=\Theta_- H-\Lambda_-
\hat{d}+\Upsilon_-\hat{c}$
(here $\Theta_\pm, \Lambda_\pm$ and $\Upsilon_\pm$ are all chiral scalar
superfields);
$U_<$ and $V_<$ are chiral superfields belonging to ${\rm exp}\hat{\cal G}_<$,
and $U_>$ and $V_>$ are chiral superfields belonging to
${\rm exp}\hat{\cal G}_>$.
Then the constraints (\ref{constraint2}) tells us that
\begin{equation}
V^{-1}_<D_-V_<=e^{{\rm ad}K_-}M_<, ~~~~~~D_+U_>U_>^{-1}=e^{-{\rm ad}K_+}
\bar{M}_>\label{constraint3}
\end{equation}
We saw in the previous section that we were able to gain the super-CAL
from the superspace WZNW reduction if we set
$\mu^1_<\mu^1_>=-1=\mu^2_<\mu^2_>$. Without loss of generality we can assume
$\mu^1_<=1=\mu^2_<$ and $\mu^1_>=-1=\mu^2_>$. Then (\ref{constraint3}) becomes
\begin{equation}
V^{-1}_<D_-V_<=e^{{\rm ad}K_-}{\cal E}_-, ~~~~~~D_+U_>U_>^{-1}=-e^{-{\rm
ad}K_+}
{\cal E}_+
\end{equation}
Moreover we have
$D_+V_<=0=D_-U_>$, $D_-K_+=0=D_+K_-$

Now calculating the quantity
$<\lambda^{(r)}_{\rm max}|G|\lambda^{(r)}_{\rm max}>$, we obtain, by
(\ref{1}),
\begin{equation}
e^{-2\lambda^{(r)}_{\rm max}(\Psi)}=<\lambda^{(r)}_{\rm max}|
e^{K_+}U_>V_<e^{K_-}|\lambda^{(r)}_{\rm max}>\label{solution2}
\end{equation}
which coincides with (\ref{reconstruction2}).

\section{Classical $r$-Matrices and Exchange Algebra}
The Poisson bracket structure in our theory is introduced
through the following positions: let
us denote with $\pi_\varphi \equiv \partial_t\varphi$
be the conjugate momentum of the bosonic field $\varphi$
($\varphi \equiv \phi, \eta , \xi$). The
non-vanishing Poisson brackets at equal time are given by
\begin{eqnarray}
\{\pi_\varphi (x,t), \varphi (y,t)\}&=& \gamma \delta (x-y)
\label{Poi1}
\end{eqnarray}
and for the fermionic component fields by
\begin{eqnarray}
\{\psi_{\pm} (x,t),\psi_{\pm} (y,t)\}&=& \mp{\textstyle {i\over 2}} \gamma
\delta
(x-y)\nonumber\\
\{\alpha_{\pm} (x,t), \beta_{\pm} (y,t)\} &=& \mp
{\textstyle{ i\gamma\over 2}} \delta (x-y)
\label{Poi2}
\end{eqnarray}
It is useful to make use of a manifestly supersymmetric
formalism. Therefore we introduce a "supercoordinate"
$X$ and a "supertime" $T$ through
\begin{eqnarray}
X&\equiv&
x={\textstyle{1\over 2}}(z_+-z_-),\quad
\theta_1= {\textstyle{1\over 2}} (\theta_++\theta_-)\nonumber\\
T&\equiv& t={\textstyle {1\over 2}} (z_+-z_-),\quad
\theta_0= {\textstyle {1\over 2}}
(\theta_--\theta_+)\nonumber
\end{eqnarray}
The integration over the Grassman variable $\theta$
 is normalized to give $\int d\theta\cdot\theta =-1$.
A superdelta function is given by
$
\delta (X,X')=(\theta_1-\theta_1')\delta (x-x')
$;
the supersymmetric step function $\vartheta (X,X')$ satisfies  $
(D_++D_-)'\vartheta (X,X')=-i\delta (X,X')$ and is given by
$
\vartheta (X,X') = \vartheta (x-x') +i\theta_1\theta_1'\delta (x-x')
$.\\
Let us introduce the conjugate supermomenta of the superfields
$\Phi , \Lambda ,\Xi$
through the positions
\begin{eqnarray}
\Pi_\Psi &\equiv &(D_+-D_-)\Psi
\end{eqnarray}
(here $\Psi\equiv \Phi ,\Lambda, \Xi$).
As a consequence of the relations (\ref{Poi1},\ref{Poi2})
we get that the superfields and their supermomenta
satisfy the following Poisson brackets
at equal "supertime":
\begin{eqnarray}
\{ \Pi_\Psi (X,T), \Psi (Y,T)\}&=& i\gamma \delta (X,Y)
\end{eqnarray}
which look the same as for the ordinary CAL theory with the replacement
of the prefix super in front of everything.\\
The integrability property of our theory is implied by the relation
for the connection ${\cal L}={\cal L}_+ +{\cal L}_-$
\begin{eqnarray}
\relax \{{\cal L}_1 (X,T), {\cal L}_2 (Y,T)\}&=& [
r_{12}^{\pm} , {\cal L}_1 (X,T) +{\cal L}_2 (Y,T) ] \delta (X,Y)
\end{eqnarray}
By definition ${\cal L}_1\equiv {\cal L}\otimes 1, {\cal L}_2\equiv 1\otimes
{\cal L}$.
The $r$-matrices $r^{\pm}$ are defined in terms of the generators
of the $\hat{osp(2|2)}^{(2)}$ algebra and are given by
\begin{eqnarray}
\relax r^+=&&{\textstyle {i\gamma\over 2}}\big[{\textstyle {1\over 2}}
H\otimes H + ({\hat c}\otimes {\hat d} + {\hat d}\otimes {\hat c})+
{\textstyle {\lambda^2\over \mu^2-\lambda^2}} H\otimes H
-{\textstyle {\lambda\over \mu}}(1+
{\textstyle {\lambda^2\over \mu^2-\lambda^2}}) H'\otimes H'+\nonumber\\
&&2(1+{\textstyle {\lambda^2\over \mu^2-\lambda^2}}) E_+\otimes E_--
(1+{\textstyle {\lambda^2\over \mu^2-\lambda^2}}) F_{\alpha_1}\otimes F_{-
\alpha_1}
-
(1+{\textstyle {\lambda^2\over \mu^2-\lambda^2}}) F_{\alpha_2}\otimes F_
{-\alpha_2}+\nonumber\\
&&
2{\textstyle {\lambda^2\over \mu^2-\lambda^2}} E_-\otimes E_++
{\textstyle {\lambda^2\over \mu^2-\lambda^2}} F_{-\alpha_1}\otimes
F_{\alpha_1}+
{\textstyle {\lambda^2\over \mu^2-\lambda^2}} F_{-\alpha_2}
\otimes F_{\alpha_2}\big]
\end{eqnarray}
for $|\lambda | < |\mu | $ and
\begin{eqnarray}
\relax r^-=&&{\textstyle {-i\gamma \over 2}} \big[{\textstyle {1\over 2}}
H\otimes H + ({\hat c}\otimes {\hat d} + {\hat d}\otimes {\hat c})+
{\textstyle {\mu^2\over \lambda^2-\mu^2}} H\otimes H
-{\textstyle {\mu\over \lambda}}(1+
{\textstyle {\mu^2\over \lambda^2-\mu^2}}) H'\otimes H'+\nonumber\\
&&2(1+{\textstyle {\mu^2\over \lambda^2-\mu^2}}) E_-\otimes E_++
(1+{\textstyle {\mu^2\over \lambda^2-\mu^2}}) F_{-\alpha_1}\otimes F_{
\alpha_1}
+
(1+{\textstyle {\mu^2\over \lambda^2-\mu^2}}) F_{-\alpha_2}\otimes F_
{\alpha_2}+\nonumber\\
&&
2{\textstyle {\mu^2\over \lambda^2-\mu^2}} E_+\otimes E_--
{\textstyle {\mu^2\over \lambda^2-\mu^2}} F_{\alpha_1}\otimes F_{-\alpha_1}-
{\textstyle {\mu^2\over \lambda^2-\mu^2}} F_{\alpha_2}
\otimes F_{-\alpha_2}\big]
\end{eqnarray}
for $|\mu | < |\lambda | $.\\
The $r^{\pm}$ matrices satisfy the classical Yang-Baxter equations
\begin{eqnarray}
\relax [r_{12},r_{13}]+[r_{12},r_{23}]+[r_{13},r_{23}] &=&0
\end{eqnarray}

The exchange algebra which summarizes all the algebraic structure underlying
the
theory can be computed with the method explained in \cite{OBabelon}.
The Poisson brackets of the vectors $\xi^{(r)},~\overline{\xi}^{(r)}$ is
\begin{eqnarray}
\{\xi^{(r)}(X)^\otimes_,{\xi}^{(r')}(Y)\}&=&\xi^{(r)}(X)\otimes{\xi}^{(r')}(Y)
[\vartheta(X,Y) r^++\vartheta (Y,X) r^-]\nonumber\\
\{~\overline{\xi}^{(r)}(X)^\otimes_, ~\overline{\xi}^{(r')}(Y)\}&=&
[\vartheta (X,Y)r^-+\vartheta (Y,X) r^+]~\overline{\xi}^{(r)}(X)\otimes
{}~\overline{\xi}^{(r')}(Y)\nonumber\\
\{\xi^{(r)}(X)^\otimes_,~\overline{\xi}^{(r')}(Y)\}&=&-\xi^{(r)}(X)\otimes
1\cdot
r^-\cdot 1\otimes ~\overline{\xi}^{(r')}(Y)\nonumber\\
\{~\overline{\xi}^{(r)}(X)^\otimes_, \xi^{(r')}(Y)\}&=&-1\otimes
{\xi}^{(r')}(Y)\cdot r^+\cdot ~\overline{\xi}^{(r)}(X)\otimes 1
\label{exchangealgebra}
\end{eqnarray}
These Poisson brackets look formally similar as the Poisson brackets of
the CAL theory and even of the Liouville theory. We stress the fact that
however
the $r$-matrices are different and that now we made use of a superfield
notation. If expressed in terms of component fields the above exchange
algebra is more singular than the corresponding algebra for the purely
bosonic theory since the Poisson bracket between fermionic fields develops
a delta function \cite{Arvis}.

\section{Concluding Remarks}
We have proposed an integrable and superconformally invariant model based on
the twisted affine KM superalgebra $\hat{osp(2|2)}^{(2)}$, that we call the
super-CAL model. This model is a manifestly $N=1$ supersymmetric
generalization of the purely
bosonic CAL theory recently proposed by BB\cite{BB1}.

The generalization to the twisted (or untwisted) affine KM superalgebras
which give rise to superconformal affine Toda theories
is possible but the analysis is expected to be more involved. The
generalization of the above manifest superspace formalism to the case with
arbitrary $(p,q)$ world-sheet supersymmetry is very interesting but will
not be straightforward.
The Coulomb gas realization of the exchange algebra (\ref{exchangealgebra})
can be provided with the methods developed in \cite{BBT}. In forthcoming
papers we will analyze the quantization of our model.
\\
\\

The authors are grateful to O. Babelon, L. Bonora,
J. Distler, R. Flume,
G. von Gehlen, W. R\"uhl, P. Schaller and M. Scheunert
for discussions.
The authors wish also thank the referee for interesting
suggestions and remarks which lead the paper to its present form.
F.T. has been financially supported by the Deutsche Forschungsgemeinschaft
(DFG) and
Y.Z.Z. by the Alexander von Humboldt Foundation (AvH).

\end{document}